\documentclass[showpacs,showkeys,prl,twocolumn,aps]{revtex4}
\usepackage{bm}
\usepackage{graphicx}
\usepackage{psfrag}
\begin{document}
\title{Riemannian geometry of irrotational vortex acoustics}
\author{Uwe R. Fischer}
\email{ufischer@uiuc.edu}
\affiliation{Department of Physics, 
University of Illinois at Urbana-Champaign\\
1110 West Green Street, Urbana, Illinois 61801--3080}
\author{Matt Visser}
\email{visser@kiwi.wustl.edu}
\affiliation{Physics Department, Washington University\\  
Saint Louis, Missouri 63130--4899\\
\ \ \ } 
\date{10 October 2001; Revised 1 February 2002; \LaTeX-ed \today}
\bigskip
\begin{abstract}
We consider acoustic propagation in an irrotational vortex, using the
technical machinery of differential geometry to investigate the
``acoustic geometry'' that is probed by the sound waves.  The acoustic
space-time curvature of a constant circulation hydrodynamical vortex
leads to deflection of phonons at appreciable distances from the
vortex core.  The scattering angle for phonon rays is shown to be
quadratic in the small quantity $\Gamma/(2\pi cb)$, where $\Gamma$ is
the vortex circulation, $c$ the speed of sound, and $b$ the impact
parameter.
\end{abstract}
\maketitle
\newcommand{\Painleve}{Painlev\'e}
\newcommand{\Iordanskii}{Iordanski\v\i}
\newcommand{\Pitaevskii}{Pitaevski\v\i}
\def\Barcelo{Barcel\'o}
\def\be{\begin{equation}}
\def\ee{\end{equation}}
\def\d{{\mathrm{d}}}
\def\half{{1\over2}}

\emph{Introduction:} 
The last few years have seen considerable interest in the development
of analog models of and for general
relativity~\cite{unruh,visser,GrishaPhysicsReports}.  These analog
models provide a two-way street: sometimes they illuminate aspects of
general relativity and sometimes the machinery of differential
geometry can be used to illuminate aspects of the analog model.  In
this article, we use mathematical methods developed in the framework
of differential geometry to study acoustic propagation in an
irrotational vortex. We analyze the ``acoustic'' space-time geometry
created by the vortex in terms of its metric tensor, Killing vectors,
and geodesics.  The medium is assumed to be ``almost incompressible'',
by which we mean that we take the background density and the speed of
sound relative to the medium to be constant, and focus on the effects
due to motion of the medium. If the flow and its perturbations are
irrotational there is a rigorous theorem to the effect that sound can
be described by a curved-space scalar wave equation using an effective
acoustic geometry, this theorem being derived by using the Euler and
continuity equations of classical hydrodynamics~\cite{unruh,visser}.
(Even if distributed vorticity is present, which is not the case
considered here, then in the eikonal approximation there are rigorous
theorems to the effect that sound ray propagation can be described by
curved-space geodesics in the same ``effective acoustic
geometry''~\cite{barcelo}.)

Using the acoustic space-time approach, we shall show that the
quasi-classical scattering process of phonons by the vortex leads to a
scattering angle quadratic in the small quantity $\Gamma/(2\pi cb)$,
where $\Gamma$ is the vortex circulation, $c$ the speed of sound, and
$b$ the impact parameter of the phonon trajectory relative to the
vortex. Due to the fact that the lowest order scattering angle is
quadratic in $\Gamma/(2\pi cb)$ and not linear, there is (up to third
order) no net modification of the transverse force exerted by a phonon
beam directed towards the vortex.  The effect of the acoustic
space-time curvature is that the vortex acts as a converging lens on
quasiclassically moving phonons.

\emph{Riemannian geometry of vortex acoustics:} 
For an irrotational vortex, the velocity profile is given in terms of
the circulation $\Gamma\equiv\oint \vec v \cdot \d\vec x\equiv
2\pi\,\gamma$ by
\begin{equation}
\vec v  = {\Gamma\over2\pi r} \, \hat\theta =  {\gamma\over r} \, \hat\theta.
\label{E:profile}
\end{equation}
If we adopt cylindrical coordinates $r$, $\theta$, and $z$, so that
the spatial metric is
\begin{equation}
g_{ij} = 
\left[
\begin{array}{ccc}
1&0&0\\
0&r^2&0\\
0&0&1
\end{array}
\right],
\end{equation}
the covariant and contravariant components of the velocity are
$v_\theta = \gamma; \, v^\theta = {\gamma / r^2}$.  Now go to a
four-dimensional description in terms of the coordinates $t$, $r$,
$\theta$, and $z$. Then, in terms of the speed of sound $c$, the
(3+1)--dimensional acoustic geometry of the vortex is described by the
space-time metric~\cite{unruh,visser}
\begin{equation}
g_{\mu\nu}^{\mathrm{vortex}} = 
\left[
\begin{array}{ccccc}
-c^2+ \left({\gamma/r}\right)^2& & 0& -\gamma &0\\
0& &1&0&0\\
-\gamma& &0&r^2&0\\
0& &0&0&1
\end{array}
\right].
\label{E:vortex}
\end{equation}
The scalar velocity potential of phonons lives in the curved
space-time world given by $g_{\mu\nu}^{\mathrm{vortex}}$.  This form
of the metric is often referred to as the {}\Painleve--Gullstrand
form~\cite{visser,PG}.  It is important to note that this vortex
metric is {\emph{not}} identical to the metric of a massless spinning
cosmic string, a solution of the Einstein equations with cylindrical
symmetry. That metric corresponds to
\renewcommand{\arraystretch}{1.0}
\begin{equation}
g_{\mu\nu}^{\mathrm{string}} = 
\left[
\begin{array}{ccccc}
-c^2 & & 0& -{\gamma} &0\\
0& &1&0&0\\
-{\gamma}& &0&r^2-\gamma^2/c^2 &0\\
0& &0&0&1
\end{array}
\right].
\label{E:string}
\end{equation}
\renewcommand{\arraystretch}{1.0}
The two metrics (\ref{E:vortex}) and (\ref{E:string}), vortex and
cosmic string, agree asymptotically at large $r$,
\begin{equation}
g_{\mu\nu}^{\mathrm{vortex}} = g_{\mu\nu}^{\mathrm{string}}
\, \left[ 1 + O(r_c^2/r^2) \right].
\end{equation}
Here we have defined the ``core radius''
\begin{equation}
r_c = {|\gamma|\over c}.
\label{E:core}
\end{equation}
This denotes the radius at which the flow goes supersonic. The
assumption of incompressibility as well as the profile
(\ref{E:profile}) certainly break down at this radius, but the actual
dimensions of the vortex core (for a superfluid vortex being set by
the coherence length $\xi_0$) are often larger than $r_c$.  The two
metrics are markedly different at intermediate values of $r$.  In
particular we shall see below that they differ significantly well
outside the vortex core radius $r_c$.  The metric of the spinning
cosmic string is everywhere flat (the space-time curvature is
identically zero).  In contrast, even at intermediate distances, the
acoustic geometry of the vortex is not flat (the space-time Riemann
tensor is not zero), and there are significant effects on the
propagation of null geodesics (sound rays) at values of $r$ well
outside the vortex core.  That there is a region of vanishing
classical deflection outside a vortex or a magnetic flux tube, like
conventionally assumed in the standard formulation of the
Aharonov--Bohm problem (see,~\emph{e.g.},~\cite{Shelankov,BerryAB}),
is thus only {\em asymptotically} true for a hydrodynamical vortex,
even if the generalized vorticity (hydrodynamical vorticity and/or
magnetic flux) vanishes everywhere outside the vortex core.  It is
hence not {\em a priori} clear that computations of the {\Iordanskii}
force based on the spinning string metric (the analog-gravitational
Aharonov--Bohm effect~\cite{grishaiord,mstone}), give the full force
exerted by a phonon beam on an actual hydrodynamical vortex.

It is straightforward to compute the Ricci curvature scalar
corresponding to the metric (\ref{E:vortex}),
\begin{equation}
R^{\mathrm{vortex}} = {2\gamma^2\over c^2\, r^4} = {2\,r_c^2\over r^4}.
\label{E:RicciScalar}
\end{equation}
The curvature of the space-time experienced by the quasiparticles is
thus significant at distances which can be well outside the core
domain.  That the flow be considered well outside the core domain is
required by the relation (\ref{E:core}), which states that at a
distance $r_c$ from the rotation axis the velocity around the core
equals the speed of sound, so that compressibility is no longer
negligible.

A simple criterion for the curvature of the effective space-time to be
significant for the motion of the phonon around the vortex at a given
distance $r$ is how the phonon wavevector magnitude $k$ compares with the
inverse space-time curvature radius at that distance. The wavevector
magnitude is then to be compared with
\begin{equation}
k_c (r) = 2\pi \sqrt R = \frac{2\pi\sqrt 2 }{r_c} \frac1{(r/r_c)^2}\,. 
\end{equation}
Well outside the core, $k_c r_c \ll 1$.  The phonon ``sees'' the
curvature of the space-time if $k$ exceeds $k_c(r)$,
and behaves as a particle moving on a geodesic in that space-time.
Vice versa, the topological structure of the acoustic space-time
related to the \Iordanskii\, force (the Aharonov-Bohm effect in the
space-time of the spinning string) dominates the behavior of the
phonons if $k\ll k_c(r)$.

For a specific physical example of an irrotational vortex
geometry, in the dilute limit of a Bose-condensed atomic vapor (BEC) 
\cite{PitaTony}, we can relate the circulation $\Gamma$, the coherence
length $\xi_0$, and the speed of sound by using the relations $\xi_0^2
=\hbar^2/(2mgn)$ and $c=\sqrt{gn/m}$. They combine to give the
relation
\begin{equation} 
2\pi \,r_c \, c =  2\pi \, \sqrt2 \,c \, \xi_0 = |\Gamma| = 2\pi\,|\gamma |
\qquad ({\rm BEC})
\label{E:circulation}
\end{equation} 
for a singly quantized vortex with $\Gamma = h/m$, where $g$ is the
strength of interaction related to the $s$-wave scattering length $a$
by $g=4\pi \hbar^2 a/m$.  The ``acoustic'' core size is thus in this
example of the same order as the actual (quantum-mechanical) core size
of varying density.  The hydrodynamic circulation, calculated with the
speed of sound along a core circumference with radius $r_c =
\sqrt 2 \xi_0$, equals the quantum of circulation in the dilute
gas limit. This relation qualitatively also holds in the dense,
strongly correlated superfluid helium II (superfluid $^4\!$He), where
$\xi_0$ is of order the atomic size.   The
relation (\ref{E:circulation}) yields for the curvature scalar in
(\ref{E:RicciScalar})
\begin{equation}
R^{\mathrm{vortex}} =\frac{4\,\xi_0^2}{r^4}. \qquad\qquad ({\rm BEC})
\label{Rxi0r}
\end{equation}

Curvature of space-time implies that particle worldlines on geodesics
in that space-time deviate from being initially parallel after some
proper distance of travel along the geodesic.  From the general
space-time interval of the \Painleve---Gullstrand metric in the
form~\cite{visser}
\begin{equation}
ds^2 = -dt^2 + \delta_{ij} (dx^i - v^i dt) (dx^j - v^j dt) \,,
\end{equation}
it is apparent that a constant time slice of the effective space-time
of quasiparticles is just ordinary Euclidean space. Hence spatial
distances measured on constant time slices in the effective space-time
are identical to distances in the Newtonian lab world, and a real
{force} is acting upon the phonon. It is, however, to be stressed that
the actual motion of the phonon in the lab world is described
correctly by the motion in the full effective space-time, that is, the
phonon is {\em not} just a (Lagrangian) particle dragged along by the
flow, if that flow is inhomogeneous.

\emph{Phonon motion in acoustic geometry:}  
To explicitly see how the vortex flow affects acoustic propagation,
and thereby get a handle on how acoustic influences can affect the
vortex, we use the eikonal approximation and consider {\emph{phonons}}
instead of sound waves. Phonons then follow null geodesics in the
acoustic geometry~\cite{visser}.  Associated with space-time
symmetries are so-called Killing vectors
\cite{MTW}, along which the metric is invariant.     
For both geometries (vortex and spinning string) 
there are three such Killing vectors,
corresponding to translations in the $t$, $\theta$, and $z$ directions:
\begin{eqnarray}
(K_1)^\mu = (1,0,0,0); \qquad (K_2)^\mu = (0,0,1,0); \nonumber\\
\hbox{and} \qquad (K_3)^\mu = (0,0,0,1).
\end{eqnarray}
This leads to three conserved quantities along each
geodesic~\cite{MTW}
\begin{equation}
(K_A)^\mu \, g_{\mu\nu} \, {\d x^\nu\over\d\lambda} = -\tilde k_A.
\end{equation}
Here $\lambda$ is some arbitrary affine parameter for the
geodesic~\cite{MTW}. We are interested in null geodesics which
represent the paths of sound rays in the acoustic geometry. From these
three conservation laws we see that for the vortex geometry
\begin{equation}
\left[-c^2+ \left({\gamma\over r}\right)^2\right] {\d t\over\d\lambda} 
- {\gamma} {\d \theta\over\d\lambda} 
= -\tilde k_1;
\end{equation}
\begin{equation}
- \gamma {\d t\over\d\lambda}  + r^2 {\d \theta\over\d\lambda} 
= -\tilde k_2;
\end{equation}
and
\begin{equation}
{\d z\over\d\lambda} 
= -\tilde k_3.
\end{equation}
By elimination between the first two equations
\begin{equation}
{\d t\over\d\lambda} = \left(\tilde k_1 + {\tilde k_2\,\gamma\over r^2}\right) c^{-2}.
\end{equation}
We are furthermore particularly interested in sound rays that come in
from infinity, so without loss of generality it is possible to
re-scale $\lambda$ to choose $\tilde k_1 = c^2$, and to then define
$\tilde k_2 = c^2 \, k_2$ and $\tilde k_3 = - k_3$. Using the new
affine parameter, a brief computation yields
\begin{equation}
{\d z\over\d t} = 
{k_3 \over
1+ {k_2\,\gamma/ r^2}},
\label{E:z}
\end{equation}
and
\begin{equation}
{\d \theta\over\d t} =  {\gamma\over r^2} -
{k_2 \,c^2/r^2
\over
1 +  {k_2\,\gamma/ r^2}}.
\label{E:theta}
\end{equation}
To now calculate $\d r/\d t$ we use the fact that the sound rays are
null curves so that
\begin{equation}
g_{\mu\nu} {\d x^\mu\over\d t} {\d x^\nu\over \d t} = 0.
\end{equation}%
Therefore
\begin{eqnarray}
&&\left[-c^2+ \left({\gamma\over r}\right)^2\right]
- 2  {\gamma} \, {\d \theta\over\d t} 
\nonumber
\\
&&
\qquad
+ \left(  {\d r\over\d t} \right)^2
+ r^2 \left(  {\d \theta\over\d t} \right)^2
+\left(  {\d z\over\d t} \right)^2 =0\,,
\end{eqnarray}
leading to, by substituting  (\ref{E:theta}) and (\ref{E:z}), 
\begin{equation}
{\d r\over\d t}=
\sqrt{
c^2- \left( 
{k_2 \,c^2/r
\over
1 +  {k_2\,\gamma/ r^2}}  
\right)^2
- 
\left( 
{k_3 \over
1 + {k_2\,\gamma/ r^2}} 
\right)^2
}.
\label{E:r}
\end{equation}
The three equations (\ref{E:z}) (\ref{E:theta}), and (\ref{E:r})
completely specify the path of the sound ray in terms of the time
parameter $t$ and the two nontrivial constants of the motion $k_2$ and
$k_3$. These have the physical interpretation that $k_3= v^z_\infty
=\beta c <c $, while in terms of the impact parameter $b$
\be
k_2 = {\gamma - c\,b \sqrt{1-\beta^2} \over c^2}. 
\ee
The radial motion is more usefully recast as an ``energy equation''
\be
\half\left( {\d r\over\d t}\right)^2 + V(r) = \half c^2,
\ee
with the ``potential''
\be
V(r) =  \half c^2 
\,{ (k_2 \,c /r)^2 + \beta^2 \over (1 +  {k_2\,\gamma/ r^2})^2}\,.
\ee
The form of this potential is sketched in figure 1 (assuming
$k_2\,\gamma>0$ and setting $\beta=0$).  Note that at large distances
it has the standard centrifugal barrier proportional to $1/r^2$, while
at short distances (however, already inside the ``acoustic'' core) it
falls quadratically to zero.
\begin{figure}[htbp]
\psfrag{r}{\large $r/r_c$}
\psfrag{V}{\large $V(r)/c^2$}
\vbox{
\hfil
\scalebox{0.85}{\includegraphics{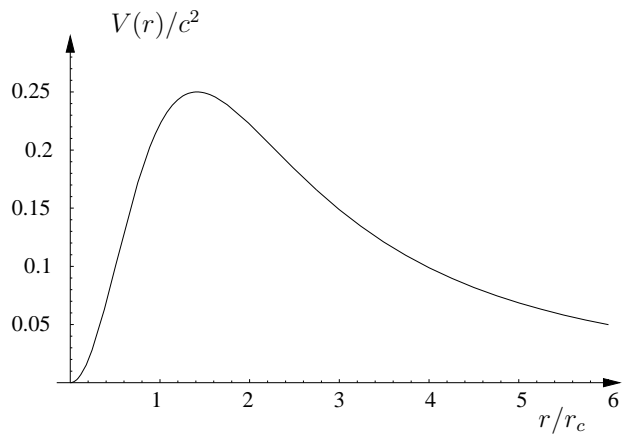}}
\hfil
}
\bigskip
\caption{
{The effective potential of phonon motion in the vortex space-time,
with $b=-r_c$ and $k_3 = v_z^\infty = 0$, $r_c=c=1$.}
}
\label{F:sketch}
\end{figure}

\begin{figure}[htbp]
\vbox{
\hfil
\scalebox{0.32}{\includegraphics{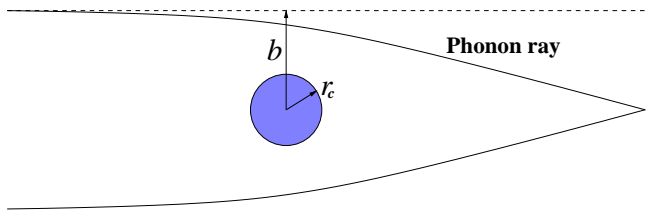}}
\hfil
}
\bigskip
\caption{
{Deflection of phonons by the vortex, which acts as a converging lens.}
}
\label{F:lens}
\end{figure}

The general deflection angle of the ray as a function of $k_2$ and
$k_3$ is obtained by integrating
\be
\Delta\theta = 2 \int_{r_{\rm turn}}^\infty {\d\theta\over\d r} 
\,\d r\,,
\label{E:deflection}
\ee
where the turning point $r_{\rm turn}$ is determined by solving
$dr/d\theta=0$. For simplicity, we set $\beta = k_3 = 0$ in the
following.  To lowest order in the small parameter $|\gamma| /cb
=r_c/b$
\be 
r_{\rm turn} = |b|
\left[1-\left(\frac {r_c}{b}\right)^2 + O\left(\frac{r_c^3}{b^3}\right) \right]
.
\ee
For $b \rightarrow \infty$, at fixed $\gamma$ and $c$ (fixed core
radius $r_c$), the deflection then evaluates from integrating
\be
{\d\theta\over\d r} = \frac{b \left(1-r_c^2/r^2\right)}{\sqrt{
(r^2 -r^2_{\rm turn})(r^2 -r_c^2)}} + O(r_c^3/b^3) 
\ee
along the ray, according to equation (\ref{E:deflection}).  The result
is to quadratic order in $r_c/b$
\begin{eqnarray}
\Delta \theta & = & \pi\, {\rm sgn} (b) \left[ 1+\frac34 
\frac{r_c^2}{b^2} + O(r_c^3/b^3) \right]. 
\label{E:angle}
\end{eqnarray}
The limit of zero classical force corresponds to the limit of
vanishing core size $r_c$ ($c\rightarrow \infty$ and/or $\gamma
\rightarrow 0$), as may be inferred from the vanishing of the
curvature scalar (\ref{E:RicciScalar}). This is directly seen in the
vanishing, to quadratic order in $r_c$, of the classical scattering of
the ray away from a straight line caused by the acoustic space-time
curvature.  The action of the vortex on the phonons is schematically
depicted in figure \ref{F:lens}.


From the deflection angle (\ref{E:angle}), 
one can easily obtain the focal length as a
function of impact parameter. Indeed two initially parallel phonon
beams that pass by opposite sides of the vortex with impact parameter
$b$ will intersect at a distance $\ell=2f$ beyond the vortex, where
\begin{eqnarray}
f = {2 \over 3\pi} {b^3\over r_c^2} \left[1+O\left({r_c\over b}\right)\right].
\end{eqnarray}
The fact that the focal length depends on impact parameter shows that
the vortex exhibits what is known in optics as ``spherical
abberation'', which in the present context is more correctly termed
``cylindrical abberation''.

In an experiment, one should take into account that the transverse
size of a phonon beam is limited by the wavelength $\lambda$, which
has to be less than or equal to the impact parameter $b$ for the
quasiclassical phonon scattering picture to make sense.  Thus the
minimum focal length is given by 
\begin{equation}
f_{\mathrm{min}} = {2 \over 3\pi}{\lambda^3\over r_c^2} 
\left[1+O\left({r_c\over \lambda}\right)\right]. \vspace*{0.25em}
\end{equation}
In superfluid $^4\!$He, where $r_c \simeq 0.6$ {\AA}, 
the roton minimum occurs at $\lambda_{\rm roton}\simeq 3$\AA.
The phonon wavelength has to be about three times the roton wavelength,  
for an approximately linear phonon dispersion to apply, 
so that $\lambda \gtrsim 15\, r_c$, which results in $f_{\rm min}
\simeq 700\, r_c$.  
Accordingly, the focal length is bounded by $f \gtrsim$ 40 nm.

In dilute Bose--Einstein condensates, due to the Bogoliubov-type spectrum
of these systems,  
which does not display a roton minimum, the phonon dispersion is to a good
approximation linear up to $\lambda \simeq 2\pi \xi_0 = \sqrt2 \pi r_c$, hence
$f_{\rm min} \simeq 20\, r_c$.   
With $r_c\simeq  0.3\, \mu $m, this results in $f \gtrsim 6 \,\mu$m 
for Bose--Einstein condensates. The latter estimate indicates that vortical
focusing effects are potentially within the realm of experimental feasibility.

\emph{Discussion:} 
An incoming phonon of finite momentum $\bm k$ passing a singular
vortex is deflected by a classical force acting upon it.  This force
is equivalent to an acoustic space-time curvature induced in the
vicinity of the vortex by that flow. From (\ref{E:angle}), it follows
that the vortex acts as a converging lens. The fact that around a
hydrodynamical vortex there is a classical force field at appreciable
distances outside the core, entails that calculations based on the
assumption that there is no force in the vorticity free region outside
the core are potentially misleading. In particular, there are finite
distance effects associated with a hydrodynamic vortex over and above
the familiar Aharonov-Bohm effect.

\emph{Acknowledgments:}
URF acknowledges support by the {\em Deutsche 
Forschungsgemeinschaft}  (FI 690/2-1) and the ESF Programme 
``Cosmology in the Laboratory''.  MV was
supported by the U.S. Department of Energy.



\begin{thebibliography}{66}

\bibitem{unruh} 
W. G. Unruh,
Phys. Rev. Lett. {\bf 46}, 1351 (1981).

\bibitem{visser} 
M. Visser, 
Class. Quantum Grav. {\bf 15}, 1767-1791 (1998);
M. Visser, Phys. Rev. Lett. {\bf 80}, 3436 (1998).

\bibitem{GrishaPhysicsReports} G. E. Volovik,
Phys. Rep. {\bf 351}, 195-348 (2001).

\bibitem{barcelo}
C.~\Barcelo, S.~Liberati, and M.~Visser,
Class.\ Quantum\ Grav.\  {\bf 18}, 1137 (2001). 

\bibitem{PG} 
P. \Painleve, 
C. R. Hebd. 
Acad. Sci. (Paris) 
{\bf 173}, 677-680 (1921); 
A. Gullstrand, 
Arkiv. Mat. Astron. Fys. {\bf 16}, 1-15 (1922).

\bibitem{Shelankov} A. L. Shelankov, Europhys. Lett. {\bf 43}, 623 (1998). 

\bibitem{BerryAB} 
M. V. Berry, 
J. Phys. A: Math. Gen. {\bf 32}, 5627 
(1999).  

\bibitem{grishaiord} 
G. E.  Volovik, 
JETP Lett. {\bf 67}, 881-887 (1998).



\bibitem{mstone} 
M. Stone, 
Phys. Rev. B {\bf 61}, 11780-11786 (2000).  

\bibitem{PitaTony} 
A. J. Leggett, 
Rev. Mod. Phys. {\bf 73}, 307-356 (2001). 


\bibitem{MTW} 
C. W. Misner, K. S. Thorne, and J. A. Wheeler: 
{\em Gravitation}, Freeman, 1973.

\end{thebibliography}
\end{document}